\documentclass[prr,aps,twocolumn,superscriptaddress,amsmath,amssymb, longbibliography,nobibnotes,nofootinbib,]{revtex4-2}
\usepackage{graphicx} 
\usepackage{soul}

\usepackage[colorlinks=true, citecolor=red,linkcolor=blue,urlcolor=magenta]{hyperref}
\usepackage{tikz}

\DeclareMathOperator{\Tr}{Tr}
\newcommand{\nc}{\newcommand}
\nc{\n}{\textrm{n}}
\nc{\rhop}{\rho_{\textrm{prod}}}
\nc{\rhos}{\rho_{\textrm{sep}}}

\nc{\lambdam}{\lambda_\textrm{min}}
\nc{\param}{s}
\nc{\Z}{\mathcal{Z}}
\nc{\id}{\mathbb{I}}
\usepackage{xcolor}

\def\rd#1{{\color{black}#1}}

\begin{document}
\title{Long-range multipartite entanglement near measurement-induced transitions}

\author{Sebastien J. Avakian}
\affiliation{\it Department of Physics, McGill University, Montr\'eal, QC, Canada H3A 2T8}
\affiliation{\it D\'epartement de Physique, Universit\'e de Montr\'eal, Montr\'eal, QC, Canada H3C
3J7}
\author{T. Pereg-Barnea}
\affiliation{\it Department of Physics, McGill University, Montr\'eal, QC, Canada H3A 2T8}
\affiliation{\it ICFO-Institut de Ciencies Fotoniques, The Barcelona Institute of Science and Technology, Castelldefels (Barcelona) 08860, Spain}
\author{William Witczak-Krempa}
\affiliation{\it D\'epartement de Physique, Universit\'e de Montr\'eal, Montr\'eal, QC, Canada H3C
3J7}
\affiliation{\it Centre de Recherches Math\'ematiques, Universit\'e de Montr\'eal, Montr\'eal, QC, Canada  H3C 3J7}

\affiliation{\it Institut Courtois, Universit\'e de Montr\'eal, Montr\'eal, QC, Canada H2V 0B3}
\date{\today}
\begin{abstract}
Measurements profoundly impact quantum systems, and can be used to create novel states of matter out of equilibrium. 
We investigate the multipartite entanglement structure that emerges in hybrid quantum circuits involving unitaries and measurements. We describe how a balance between measurements and unitary evolution can lead to multipartite entanglement spreading to distances far greater than what is found in non-monitored systems, thus evading the usual fate of entanglement. We introduce a graphical representation based on spanning graphs that allows to infer the evolution of genuine multipartite entanglement for general subregions. We exemplify our findings on hybrid random Haar circuits that realize a 1d measurement-induced dynamical phase transition, where we find genuine 3-party entanglement at all separations.  At criticality, our data is consistent with power-law decay with a tripartite exponent strictly larger than the one of the bipartite logarithmic negativity. The 4-party case is also explored.
Finally, we discuss how our approach can provide fundamental insights regarding entanglement dynamics for a wide class of quantum circuits and architectures.
\end{abstract}
 
\maketitle

\section{Introduction}
Measurements profoundly impact quantum systems, especially their quantum entanglement. A perfect measure of a spin's component collapses its wavefunction into a pure state un-entangled with other parts of the system. Such a loss of entanglement can actually promote entanglement among the remaining degrees of freedom of the system as they become ``liberated'' from the measured spin. But the decoupling of the measured spin is only temporary, since being in a pure product state, it is very ``entanglable,'' namely it can be readily entangled with its environment. In contrast, a highly mixed (decohered) spin would not be entanglable.
We thus see that measurements can have  highly non-trivial effects on the entanglement structure in quantum many-body systems. Partial monitoring involving measurements on parts of a system can be used to reach new non-equilibrium regimes beyond what is possible in usual unitary evolution. An example is the appearance of measurement-induced dynamical phase transitions in quantum circuits with unitary and measurement layers, see Fig.~\ref{fig:1d}. It was discovered that as one increases the measurement rate, a continuous transition from a volume law of the von Neumann entanglement entropy to an area law occurs \cite{Skinner2019,Li2018,Li2019,Jian2020,Chan2019,pc1,rev}. Interestingly, experimental signatures of such a transition were observed on superconducting quantum processors with mid-circuit measurements via entanglement entropies~\cite{koh2023measurement}.
A deeper theoretical analysis of the bipartite entanglement yielded the numerical observation that the critical point has a logarithmic negativity~\cite{LN1,LN2} for two intervals that decays algebraically with separation (after ensemble averaging) \cite{Shi2021,Sang2021}. Such long-range entanglement is striking given that quantum matter at equilibrium will typically have short-ranged  entanglement, both bipartite and multipartite, owing to the fate of entanglement under general types of evolution, both in space and time \cite{Parez2024_F}. 
A striking example can be found in quantum critical ground states described by conformal field theories where the bosonic logarithmic negativity between separated subregions decays faster than any power in one~\cite{CCT12, CCT13} and higher dimensions~\cite{Parez2024_B}. 

In this work, we investigate the multipartite entanglement structure in general quantum circuits involving unitaries and measurements. We explain how a balance between measurements and unitary evolution can lead to multi-party entanglement spreading to distances far greater than what is found in non-monitored systems. We introduce a graphical representation that allows to identify the evolution of multipartite entanglement structure, and exemplify it on simple circuits. We support our analysis by computing measures of genuine multipartite entanglement (GME) for various subregions. In essence, GME is a collective form of entanglement that involves all parties. As the key example, we study 3-spin GME near the measurement-induced transition in 1d random Haar circuits, where we find GME at all separations. In particular, when we work at the critical rate $p_c$, \rd{our data is consistent with} power-law decay for both the negativity and tripartite GME, with distinct quantum critical exponents. The tripartite GME is evaluated using two separate metrics, yielding the same exponent.
Our negativity exponent is larger than previous estimates on shorter chains~\cite{Sang2021,Shi2021}, while the tripartite exponent has not been previously studied. \rd{By studying finite-size and finite-ensemble effects, we point out the limitations of the present analysis, which should help guide future investigations. } 
We further provide non-pertubative arguments for the scaling of GME measures as a function of the rate $p$, which suggests that the non-analytic $p$-scaling can be difficult to observe.
We then analyze the layer-by-layer dynamics and show how unitaries and measurements work in tandem to produce genuine multiparty entanglement in subregions. We end by introducing \emph{minimal spanning graphs} connecting the various parties, and show how these provide a framework that assists in the identification of GME in a large family of quantum circuits.

Multipartite entanglement has been studied in certain monitored quantum spin chains through the quantum Fisher information.  It was found that the latter detects the measurement-induced transition~\cite{DiFresco23,Paviglianiti23,Fazio2024}. In contrast to the current work, the Fisher information was computed for the entire chain, and could not reveal the structure of GME for subregions.  

\section{Evading the fate of entanglement}
\begin{figure}
\centering
\includegraphics[width=0.49\textwidth]{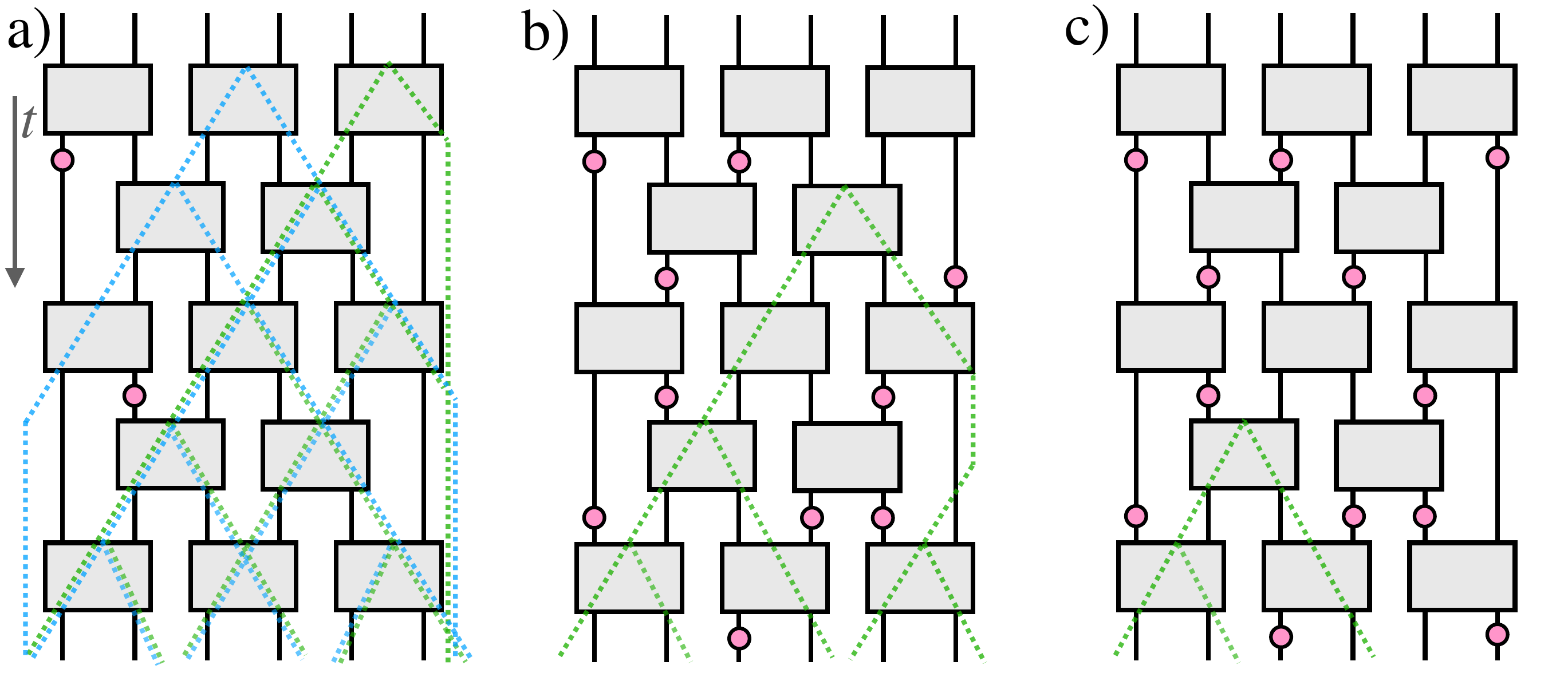} 
\caption{{\bf Circuits and entanglement spreading}. Three quantum circuits composed of two-site unitaries (grey boxes) and measurements (pink circles); the measurement rate increases left to right. 
Time flows downwards. Entanglement graphs that propagate to the final spins are shown. Different colors in (a) correspond to distinct seeds.  }
\label{fig:1d}
\end{figure}

Let us look into the time evolution of a discrete quantum circuit, a simple 1d example is shown in Fig.~\ref{fig:1d}.
The general circuits under consideration are composed of unitary operators (boxes) that act on a group of nearby sites, and projective measurements (circles). 
Our goal is to understand the evolution of multipartite entanglement within general subregions. Such a subregion $A$ is composed of $m$ groups of spins, $A_1,\ldots,A_m$, and can thus possess up to $m$-party entanglement. We will see how the right amount of measurements can allow the system to evade the typical fate of entanglement~\cite{Parez2024_F}.

Measurements tend to decrease the amount of collective entanglement in the entire system since the measured spins factorize from the rest. However, they can increase multipartite entanglement within $A$ by removing entanglement to the spins in the complement, $B$. Indeed, if a spin in $B$ is entangled with $A$, but is hit by a measurement, this entanglement will be destroyed potentially allowing for more entanglement within $A$. This follows from monogamy: a spin in $A$ that is strongly entangled with $B$ cannot maximize its entanglement with other spins in $A$. 
In contrast, if a measurement hits a spin in $A$, this spin will no longer contribute to the entanglement within $A$.

Unitaries generate the entanglement in the first place. Consider a product state of 2 spins acted upon by a unitary gate. The resulting state will generically be entangled. However, successive applications of the unitary will not increase entanglement indefinitely: decrease will eventually occur. Turning to a general subregion $A$, a degradation of entanglement will occur if various unitaries act upon the subregion without being interrupted by measurements due to scrambling. In fact, measurement-free unitary evolution will typically drive $A$ towards and into the separable space of states \cite{Parez2024_F}.
We thus see that a balance between measurements and unitaries must be achieved to allow entanglement to spread. 

\begin{figure}
\centering
\includegraphics[width=0.5\textwidth]{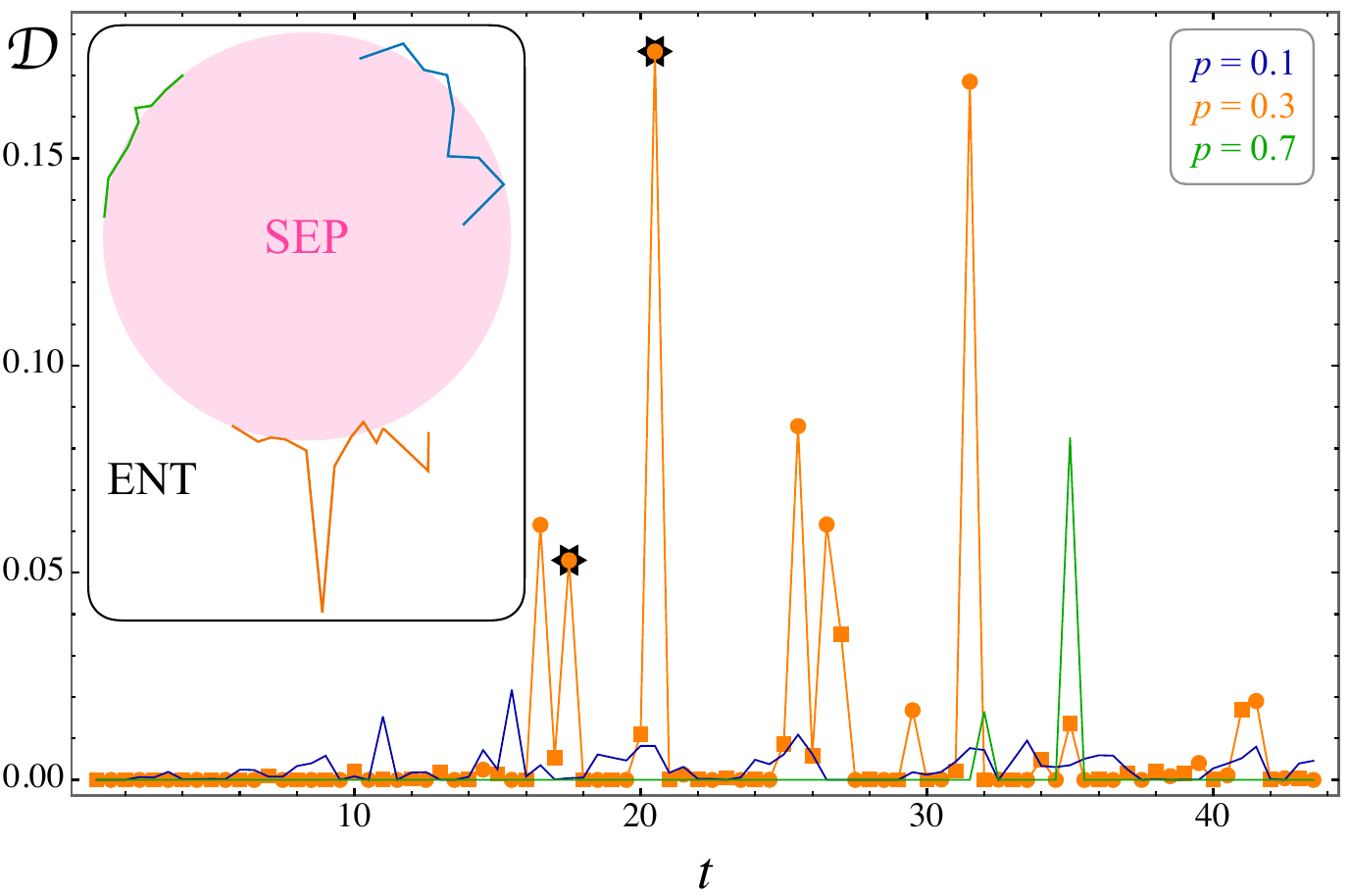}
\caption{{\bf Time evolution.} The geometric entanglement ${\cal D}$, which measures the distance to the separable set, versus discrete time for 3 spins at positions $(4,7,10)$ in a $L=14$ chain. $\mathcal D\!>\! 0$ indicates the presence of bipartite or tripartite entanglement. 
Unitaries occur at integer times $t=1,2,\dots$, while measurements at $t=3/2,5/2,\dots$. Curves are shown for three values of $p$ (see legend). For $p=0.3$, post-unitary states are marked with squares, and post-measurement ones with circles.  
Black stars denote detection of GME. {\bf Inset.} The separable set (SEP) is surrounded by entangled states (ENT). Typical time evolution paths are schematically shown for the three $p$ values. }
\label{fig:time}
\end{figure}

The inset of Fig.~\ref{fig:time} schematically illustrates these principles. We represent the space of states on subregion $A$ that have no entanglement, i.e.\ classical mixtures of product states of the form $\rho_1\otimes\cdots\otimes\rho_m$, by a disk surrounded by a sea of entangled states. We then illustrate the typical layer-by-layer time-evolution of $\rho_A(t)$ in the regimes of low (blue), intermediate (orange), and high (green) measurement rates. At low rates, scrambling limits entanglement, and the state actually penetrates within the separable continent. At high rates, the state spends most of the time on the boundary of the separable set; that is where pure product states live. Measurements prevent $\rho_A$ from penetrating the separable continent. At intermediate rates, measurements still prevent the state from penetrating too deep into the separable continent, allowing unitaries to generate substantial entanglement. Below we shall provide quantitative analysis that precisely supports this qualitative picture. 

\section{Multipartite entanglement spreading} 
We examine in more detail how multipartite entanglement spreads in quantum circuits. In particular, we want to identify subregions $A$ that possess GME, which means that the reduced density matrix $\rho_A$ is not biseparable. For example, in the case of three subregions $A=A_1A_2A_3$, a biseparable state is a mixture of states separable under some partition:
\begin{align}	\label{eq:bisep}
	\rho_{\rm bisep} = \sum_k p_k \rho_1^k\otimes\rho_{23}^k+ q_k \rho_{13}^k\otimes \rho_2^k+r_k \rho_{12}^k\otimes\rho_3^k
\end{align}
where $p_k,q_k,r_k\geq 0$, and the $\rho^k_\#$ are physical density matrices.
An elementary graphical representation will allow us to visualize the evolution of the entanglement structure. We exemplify the representation in Fig.~\ref{fig:1d}. Starting at the earliest times, one draws a cone of entanglement that arises from a unitary (a ``seed''), as long as measurements do not prevent the cone from reaching the final state. We thus first obtain the cones with the largest spreads. For a given cone, one then draws all sub-cones that reach the final state. The procedure is repeated layer-by-layer until all possible cones are identified. The end result ressembles a root system connecting the final spins, see Fig.~\ref{fig:1d}. For $m$ groups of spins to be entangled, they should belong to a connected root system. 
However, this is not sufficient since too many overlapping root systems can scramble the spins belonging to the $m$ parties. Indeed, different root systems possess different seeds, so that the spins connected by too many overlapping roots receive quantum information from independent sources, preventing them from reliably encoding the information coming from any given seed. A low rate of measurements gives many overlapping roots, with little resulting entanglement, see Fig.~\ref{fig:1d}a. In contrast, a high  measurement rate only allows short roots to grow, strongly suppressing the range of entanglement, as shown in Fig.~\ref{fig:1d}c. For intermediate rates (Fig.~\ref{fig:1d}b), one achieves a root spanning numerous sites resulting in long-range entanglement. To see which spins are entangled, one examines the sites connected by roots. 
In section \ref{sec_6}, we provide a more precise graph-based construction, and demonstrate how it can indicate the presence and strength of GME.

We can quantify the various forms of entanglement. To do so we study the circuits in Fig.~\ref{fig:1d}, and fix the unitary to be the Floquet Ising one used in Ref.~\cite{Skinner2019}. The sole source of randomness arises from the probabilistic outcomes of measurements in quantum mechanics.
In the regime of low measurement rate, Fig.~\ref{fig:1d}a, we find no long-range GME irrespective of the measurement outcomes. For instance, we detect no genuine tripartite entanglement between 3 sites when the distance between the first and last spins obeys $i_3-i_1>3$. For 4-sites, we found no GME of maximal range $i_4-i_1=5$. To reach these conclusions we used bi-separability criteria \cite{W_criterion}, the key one being
\begin{align} \label{eq:W}
 W= \max_{\rm LU}  |\rho_{18}| -  \sqrt{\rho_{22}\rho_{77}} - \sqrt{\rho_{33}\rho_{66}} - \sqrt{\rho_{44}\rho_{55}}
\end{align} 
where $\rho$ is the three-spin reduced density matrix in the canonical computational basis: $|1\rangle=|00 0\rangle, |2\rangle=|001\rangle, \ldots, |8\rangle=|111\rangle$.
$W>0$ indicates that the state cannot be written as \eqref{eq:bisep}, and thus GME exists among the 3 spins. If the RHS is non-positive a conclusion cannot be reached, and we set $W=0$. The maximisation is over all local unitary transformations on the 3-spin density matrix $\rho_{ij}$.
\rd{It turns out that $W$ is a lower bound for a true measure of GME, the genuinely multipartite concurrence (or GME concurrence), which is defined by using a convex roof extension of the bipartite concurrence (see \cite{Ising123D} and references therein).}
We employed a similar criterion for 4-spin GME~\cite{W_criterion}, $W_4$, as explained in Appendix~\ref{ap:four}. 
Moving to the intermediate measurement rate circuit in Fig.~\ref{fig:1d}b, we detect GME between sites 1,2,4 and sites 2,4,6; the former is stronger since the root system connecting the spins is shorter. 
We also find long-range 4-partite GME between sites 1,2,5,6. Finally, in the high measurement rate regime, the only detected 3-spin GME is 1,2,4 owing to the corresponding root system in Fig.~\ref{fig:1d}c. We thus see the general principles, and graphical representation of GME at work in a simple example. 

\section{Multipartite entanglement in random Haar circuits}
\begin{figure}
\centering
\includegraphics[width=0.48\textwidth]{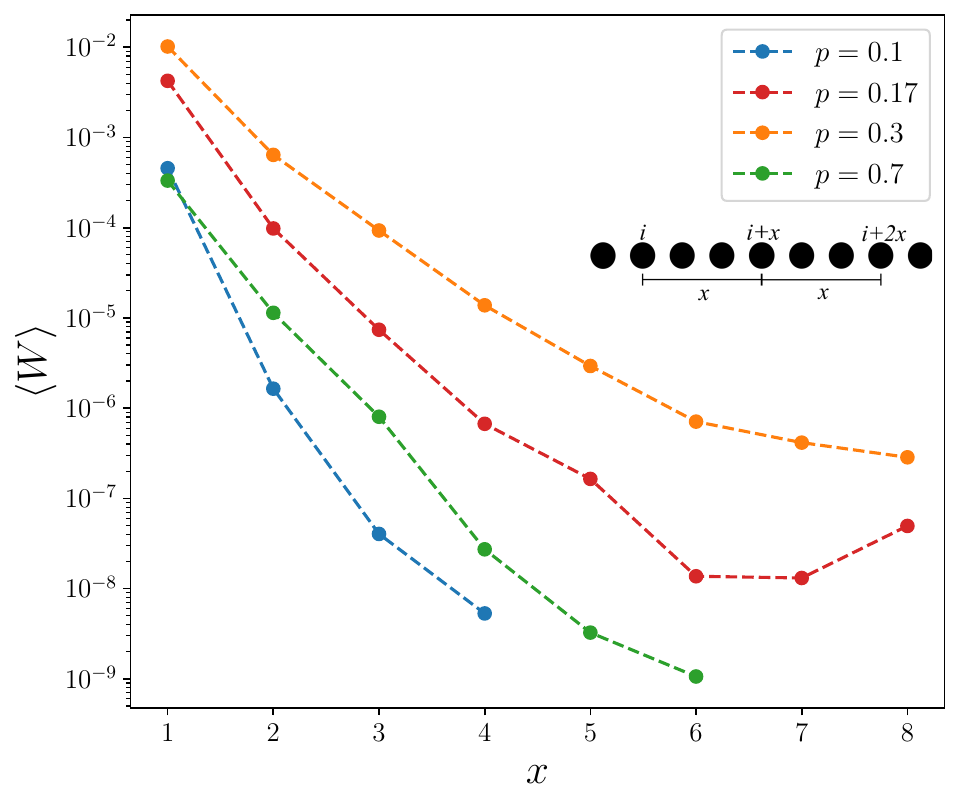}
\caption{{\bf Genuine multipartite entanglement in random Haar circuits.} Log-linear plot of the ensemble average of the criterion $W$, Eq.~\eqref{eq:W}, that detects GME among 3 spins at positions $(i,i+x,i+2x)$ in $L=18$ chains with open boundary conditions. Longer-range GME is observed for a measurement rate of $p=0.3$. 
For $p=0.1,0.3,0.7$, the ensemble averages were taken over $4\times 10^5$ realizations while for $p=0.17$ the ensemble averages are taken over $10^6$ realizations.}
\label{fig:W}
\end{figure}

We now turn to a family of circuits that realize a bona fide measurement-driven phase transition \cite{Skinner2019,Li2018,Li2019,Jian2020,Chan2019}. The circuit structure appears in Fig.~\ref{fig:1d}, and the unitaries are  chosen randomly via the Haar measure. Single-site $Z$-measurements are performed with a rate $0\leq p\leq 1$. In Fig.~\ref{fig:W} we show results for chains with $L=18$ sites at four values of $p$. We plot the biseparability criterion $W$ for subregion $A$ composed of $m=3$ spins at positions
$(i,i+x,i+2x)$; the maximal possible range being $x=8$. We have averaged over $4\times 10^5$ samples for $p=0.1,0.3,0.7$ while for $p=0.17$ we averaged over $10^6$ 
circuit realizations. The transition between volume and area-law regimes for hybrid random Haar circuits has been found to be at $p_c=0.17(1)$~\cite{pc1}. 
We see that the intermediate value $p=0.3$ shows a bigger $\langle W\rangle$ compared to $p=0.17$, as does the logarithmic negativity, see Fig.~\ref{fig:L24NPBC} and discussion below.
Among the four rates, $\langle W\rangle$ is largest for $p=0.3$, and has extends up to to $x=8$. 
In contrast, for $p=0.1$ we have only  detected events with $x\leq4$. For $p=0.17$ we have events for all $x$, where $\langle W \rangle$ is smaller than at $p=0.3$, with the caveat that at $x=8$ we have only 3 hits which skews the averaging. For $p = 0.7$, we only found events with $x\leq6$, and the average value at $x=6$ is many orders of magnitude smaller than for $p=0.3$. We note that the lack of data points up to $x=8$ for $p\neq 0.3$ is probably due to the following reasons: lack of samples, $W$ does not capture all GME, the optimisation required to get $W$ is \rd{imperfect, especially} at large $x$, where entanglement is weak. We come back to this point at the end of the paper where we introduce a powerful graphical representation that will allow us to infer the presence and strength of GME. 

To appreciate the striking nature of the above results, it is important to put these into context. Let us compare with the analogous analysis for the transverse field Ising model near its quantum critical point in 1d. 
GME between 3 adjacent spins ($x=1$) is detected by $W$, and takes its maximal value close to the critical point \cite{Giampaolo2013,Giampaolo2014,Ent_Microscopy}. However, as soon as $x>1$, GME has not been detected \cite{Giampaolo2013,Hofmann_2014,Ent_Microscopy}, and the 3-spin reduced density matrix has been shown to be biseprable~\cite{Hofmann_2014}. This means that $W$ vanishes for $x>1$, in clear contrast to what we observe in  monitored quantum circuits. An even more striking comparison can be made with the 2d quantum Ising model, where $W$ is both weaker and occupies a smaller fraction of the phase diagram compared to the 1d case; it also vanishes for non-adjacent sites \cite{Ent_Microscopy}. In contrast to such sudden death, which is generically expected in equilibrium~\cite{Parez2024_F}, we find an algebraically slow decay of genuine tripartite entanglement at the non-equilibrium quantum critical point, as discussed below.

\subsection{Scaling near criticality}
\begin{figure}
\centering
\includegraphics[width=0.48\textwidth]{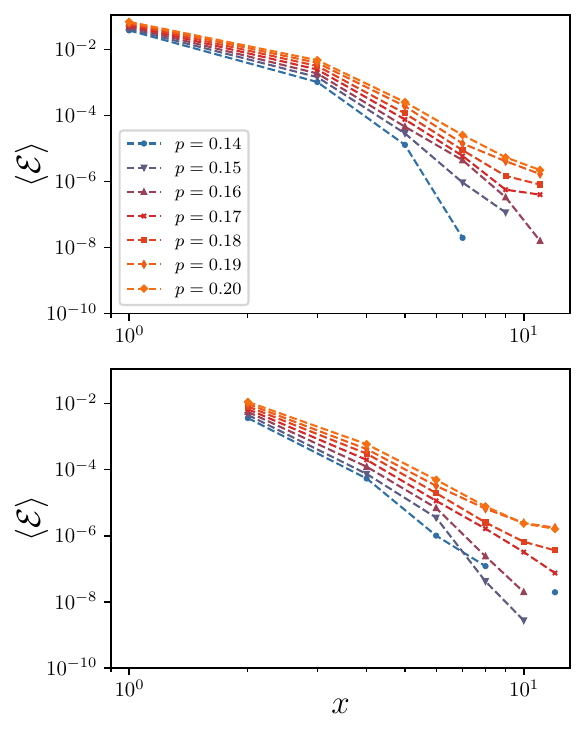}
\caption{ {\bf 2-spin logarithmic negativity for $L=24$ with PBC}. Log-log plot of the logarithmic negativity $\mathcal{E}$ for 2 spins at positions $(i,i+x)$ for $L=24$ chains with periodic boundary conditions. Due to the nature of brickwork circuits, there is an even and odd effect in $x$. The \rd{top} panel includes only the odd separations $x=1,3,5,7,9,11$ while the \rd{bottom} panel includes the even separations $x=2,4,6,8,10,12$. The number of realizations for $p=0.14,0.15,0.16,0.17,0.18,0.19,0.20$ are respectively $3.41\times 10^4,3.37\times 10^4,3.32\times 10^4, 1.05\times10^5,3.17\times10^4,2.56\times10^4,2.73\times10^4$.}
\label{fig:L24NPBC}
\end{figure}
\begin{figure}
\centering
\includegraphics[width=0.48\textwidth]{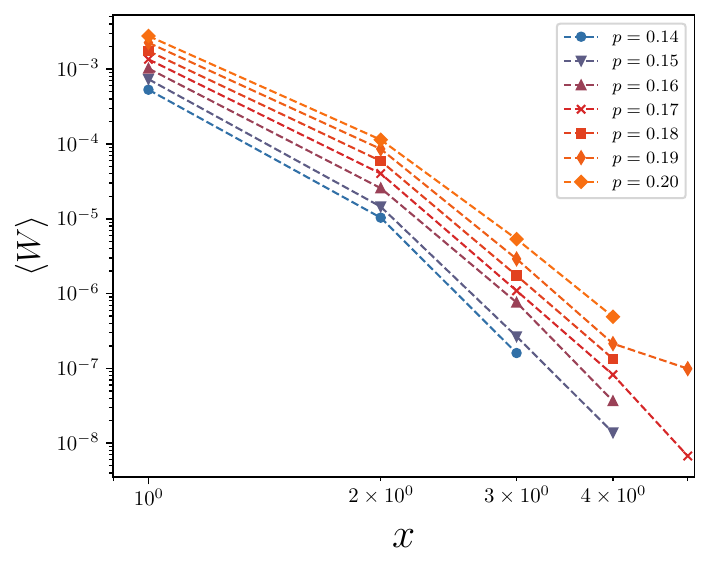}
\caption{{\bf 3-spin GME with periodic boundary conditions near $p_c$}. Log-log plot of the biseparability criterion $W$ for 3 spins at positions $(i,i+x,i+2x)$ in a $L=24$ chain with periodic boundary conditions. The number of realizations for $p=0.14,0.15,0.16,0.17,0.18,0.19,0.20$ are respectively $3.41\times 10^4,3.37\times 10^4,3.32\times 10^4, 1.05\times10^5,3.17\times10^4,2.56\times10^4,2.73\times10^4$. While the largest separation for $L=24$ is $x=8$, only events up to $x=5$ have been detected.
}
\label{fig:L24WPBC}
\end{figure}

We now examine more closely the behavior near the critical rate. 
Fig.~\ref{fig:L24WPBC} shows $\langle W\rangle$ for $p=0.14,0.15,0.16,0.17,0.18,0.19,0.20$, this time with a larger number of qubits, $L=24$, and periodic boundary conditions. We observe that $\langle W\rangle$ increases in a monotonous fashion with $p$ \rd{in the given region} for all $x$ that have positive hits.
In particular, maximum entanglement is not attained at the transition. 
To further consolidate the point,
Fig.~\ref{fig:L24NPBC} \rd{also} shows the ensemble average of the logarithmic negativity between 2 spins as a function of separation: it shows a similar $p$-dependence as $W$, i.e.\/ the maximum does not occur at the transition. 
This is reminiscent of quantum critical systems in equilibrium. The critical transverse field Ising chain shows a $W$ for 3 adjacent spins that reaches its maximum at a field value strictly greater than the critical value~\cite{Ent_Microscopy} $h_{\rm max}>h_c$. This is an exact result since the exact solution in the thermodynamic limit was used. 
We thus see that 1d hybrid circuits show similar dependence for their 3-spin GME on the tuning parameter as a spin chain in equilibrium.

Furthermore, a general argument has been given~\cite{Ent_Microscopy} for equilibrium quantum critical theories stating that generic entanglement measures  have singular behavior as a function of the non-thermal parameter. The relation (in systems where the symmetry is not spontaneously broken such as finite lattices) is that an entanglement measure $\mathcal M$ scales as 
$\mathcal M= \mathcal M_c + \alpha |h-h_c|^{\Delta_\varepsilon \nu}+\cdots$, where $h$ is the non-thermal tuning parameter, $\Delta_\varepsilon$ is the (spatial) scaling dimension of the relevant scalar that tunes the system to criticality, and $\nu$ the correlation length exponent.
This will lead $d\mathcal M/dh$ to have a divergence at the critical point if $\Delta_\varepsilon \nu<1$ (otherwise, a higher derivative will show the leading divergence). For the 1d transverse field Ising chain we have $\Delta_\varepsilon=2-1/\nu=\nu=1$, so that the relation leads to a logarithmic divergence for $W$ of 3 adjacent sites, which has been verified by using the exact solution~\cite{Ent_Microscopy}. For a system tuned near a measurement-induced transition, we also expect singular scaling of a generic entanglement measure once it has been ensemble-averaged, $\langle \mathcal M\rangle$. In order to determine the critical exponent of the divergence, one would need to identify the operator with the lowest scaling dimension in the non-unitary critical field theory that acquires an expectation value; it can differ on either side of the transition. Identifying such scaling in our numerical data proves to be too challenging owing to the need to get numerous results near $p_c$, each requiring considerable ensemble averaging and finite-size scaling.
Instead, we can attempt to estimate the scaling based on the exponents for 2d percolation, which are believed to be close to the Haar measurement-induced transition \cite{pc1}: $\nu^{\rm P}=4/3$ and $\Delta_\varepsilon^{\rm P}=2-1/\nu^{\rm P}$. \rd{By the above argument,} we would then obtain a singular contribution to a generic entanglement measure $\sim |p-p_c|^{\Delta_\varepsilon^{\rm P}\nu^{\rm P}}$ with $\Delta_\varepsilon^{\rm P}\nu^{\rm P}=5/3$, such that the second derivative $d^2\langle \mathcal M\rangle/dp^2$ would show a divergence at criticality, in contrast to the quantum Ising transition in equilibrium where the first derivative is divergent for generic measures $\mathcal M$. Such behavior would thus be even more challenging to observe in the numerics. \rd{A careful analysis of the dominant singular scaling is beyond the scope of the present paper, and would deserve further work.}

\subsection{Spatial scaling at criticality}
Let us now analyze the $x$-dependence of bipartite and tripartite measures at the critical rate $p=0.17$. Regarding the latter, we consider an additional biseparability criterion~\cite{Huber2010}:
\begin{equation}  \label{eq:I2}
    I_2(\rho)=\max _{| \Phi \rangle}\left\{\sqrt{\langle\Phi| \rho\otimes \rho \Pi|\Phi\rangle}-\sum_i \sqrt{\langle\Phi| \mathcal{P}_i^\dagger \rho\otimes \rho \mathcal{P}_i|\Phi\rangle}\right\}
\end{equation}
where the above expression involves two copies of the $m$-party state under consideration, $\rho \otimes \rho$. The maximization is over all product states in the doubled Hilbert space:
$\vert \Phi \rangle = |\varphi_1\rangle \otimes|\varphi_2\rangle \otimes \ldots \otimes|\varphi_{2m}\rangle$.
$\Pi$ is the global swap operator that exchanges the two copies, while $\mathcal{P}_i=\Pi_{A_i} \otimes 1_{\!B_i}$ is the swap operator for bipartition $A_i\vert B_i$ which exchanges part $A_i$ of the two copies. For $m=3$ single-spin parties, there are 3 bipartitions: $1|23, 12|3, 13|2$. If $I_2 > 0$, there is GME, while no conclusion can be made if $I_2 \leq 0$. We note that $I_2$ is more powerful than $W$ as it detects GME-states missed by the latter, however this comes at an increased computational cost due to the larger number of parameters in the optimization.
\rd{Interestingly, it can be shown~\cite{Ising123D} that $I_2$ is not only lower bounded by $W$, but that it also provides a stronger lower bound for the genuinely multipartite concurrence, a true measure of GME. }

\begin{figure}
\centering
\includegraphics[width=0.5\textwidth]{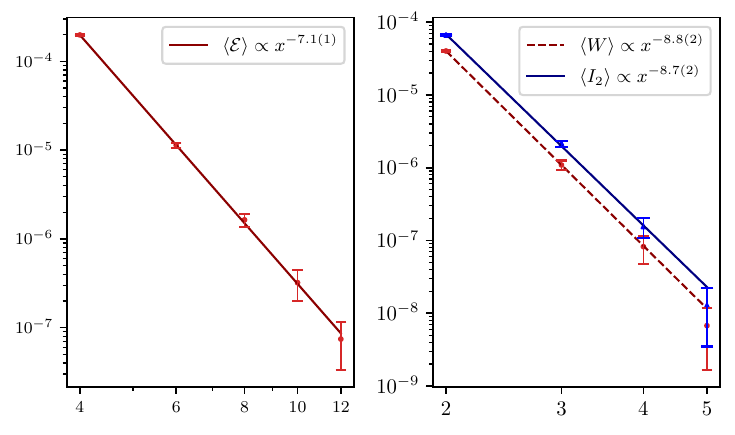}
\caption{
{\bf{Scaling of $\langle \mathcal{E}\rangle$, $\langle W \rangle$, and $\langle I_2 \rangle$ at $p=0.17$ vs $x$}}. Log-log plot of the $x$-even 2-spin logarithmic negativity (left panel), and the 3-spin $W$ and $I_2$ criterions at $p_c=0.17$ (right panel) with their corresponding error bars for $N=24$ spins with PBC. The fitting is performed without the largest $x$-value due to rare event fluctuations that skew  the ensemble average. 
}
\label{fig:powerlaw}
\end{figure}

In Fig.~\ref{fig:powerlaw}, we show power-law fits of the form $1/x^\alpha$  for $\mathcal E$, $W$, and the new measure $I_2$. We did not use the last data point to obtain the fit as it is subject to large fluctuations due to rare events. We obtain \rd{the following estimates} 
\begin{align}
\alpha_{\mathcal E}=7.1(1)\,, \quad \alpha_W=8.8(2)\,, \quad \alpha_{I_2}=8.7(2)     \end{align}
where the error bars are the standard deviation divided by the square root of the total population size, which should be understood as a lower bound for the true uncertainties, \rd{which are certainly larger (more on this below)}.
First, we note that both genuine tripartite entanglement measures decay with the same power within error bar. 
Second, tripartite entanglement decays faster than the bipartite logarithmic negativity. This is expected since having 3 spins at positions $(i,i+x,i+2x)$ being genuinely entangled is more difficult to achieve than having 2-spin entanglement. Indeed, typical bi-separable states of 3 spins will have 2-spin bipartite entanglement, yet no genuine 3-spin entanglement. Third, we can compare our negativity exponent with previous results in the literature. In Ref.~\cite{Sang2021}, a smaller chain length $L=22$ and \rd{rate} of $p=0.22$ were used to obtain $\alpha_{\mathcal E}\approx 6.2$. In Ref.~\cite{Shi2021}, with shorter $L=20$ chains and \rd{a larger value of the} rate $p=0.26$, the following estimate was obtained $\alpha_{\mathcal E}=5.46\pm0.56$. In those references,  intervals of various sizes were used in the fit, which can \rd{potentially} exacerbate finite size effects in short chains.
\rd{To get a better understanding of the latter, we have studied shorter systems: $N=16$ with PBC at $p=0.17$. The results are shown in Appendix~\ref{ap:numerics}. For the LN between two spins, a power law fit of the data as a function of $x$ yielded $\alpha_{\mathcal E}=5.3(2)$. However, given that such a fit could only be made with two points in such short chains, we also performed a fit by using the chord length, allowing a reasonable 3-point fit that gave $\alpha_{\mathcal E} =8.0(2)$. Such a result is larger than what we reported above for $N=24$. Regarding the tripartite measures, the direct fits with the distance $x$ yielded $\alpha_W=7.1(1)$ and $\alpha_{I_2}=6.8(1)$, which are smaller than the $L=24$ exponents. Just as for $L=24$, the $N=16$ tripartite results did not show the anticipated upturn at large separations necessary for chord length scaling. This is expected given that the large-$x$ hits are rare, and would require a larger ensemble of circuit realizations, which is beyond the scope of the current work.
Overall, although we see encouraging results across different sizes, the finite-size and finite-ensemble effects are important in these systems, and require a more systematic analysis. }


Interestingly, our negativity exponent $\alpha_{\mathcal E}=7.1$ is distinct from what was obtained for hybrid Clifford circuits in the two above references: $\alpha_{\mathcal E}^{\rm Cliff}\approx 6$. It would be interesting to certify whether the exponents indeed differ in both types of circuits. Our results also motivate an analysis in the continuum via the putative non-unitary conformal field theories describing such measurement-induced transitions.


\section{Time evolution}

We quantitatively study how entanglement evolves in time layer-by-layer. To obtain a more complete characterization, we shall employ the geometric entanglement \cite{Geo_Ent} 
\begin{align}
	\mathcal D = \min_{\rho_{\rm sep}} d(\rho,\rho_{\rm sep}) 
\end{align}
that measures the distance between $\rho$ and the closest state $\rho_{\rm sep}$ in the separable continent, as pictured in the inset of Fig.~\ref{fig:time}. We shall use the Frobenius norm to define the commonly used Hilbert-Schmidt distance: $d(\rho_1,\rho_2)=\sqrt{\Tr (\rho_1-\rho_2)^2}$. The strength of $\mathcal D$  is that it captures all forms of entanglement. Fig.~\ref{fig:time} shows the time-evolution of $\mathcal D$ for 3 spins at positions $(4,7,10)$, which corresponds to $x=3$, in a chain of $L=14$ sites. 
For every time step $t$, we find the separable state of 3 spins nearest to $\rho_A(t)$, $\sum_k p_k\rho_1^k\otimes\rho_2^k\otimes\rho_3^k$ with the $p_k$ forming a probability distribution. The numerical optimisation is done over $9k+(k-1)=69$ real parameters (we parametrize a mixed product state with $3\times 3$ parameters), i.e.\ we go up to $k=7$. 
We have verified numerous circuit realizations, and the ones shown represent typical behavior. We see that that $\mathcal D$ remains small both at $p=0.1$ and $p=0.7$. However, at the intermediate rate $p=0.3$, $\mathcal D$ exhibits recurring large spikes, meaning that $\rho_A$ has substantial bipartite or tripartite entanglement at these times. In fact, the highest spikes occur for post-measurement states (circles). We observe the following mechanism to build a large spike: a unitary layer (square in Fig.~\ref{fig:time}) generates a small amount of entanglement in $A$, which is then amplified by an appropriate measurement layer (circle). In all the data shown, we only detect 3-spin GME at two times (black stars), and only for $p=0.3$. We thus see that many of the high spikes are dominated by bipartite entanglement, and possess correspondingly little or no GME. 

\begin{figure}
\centering
\includegraphics[width=0.5\textwidth]{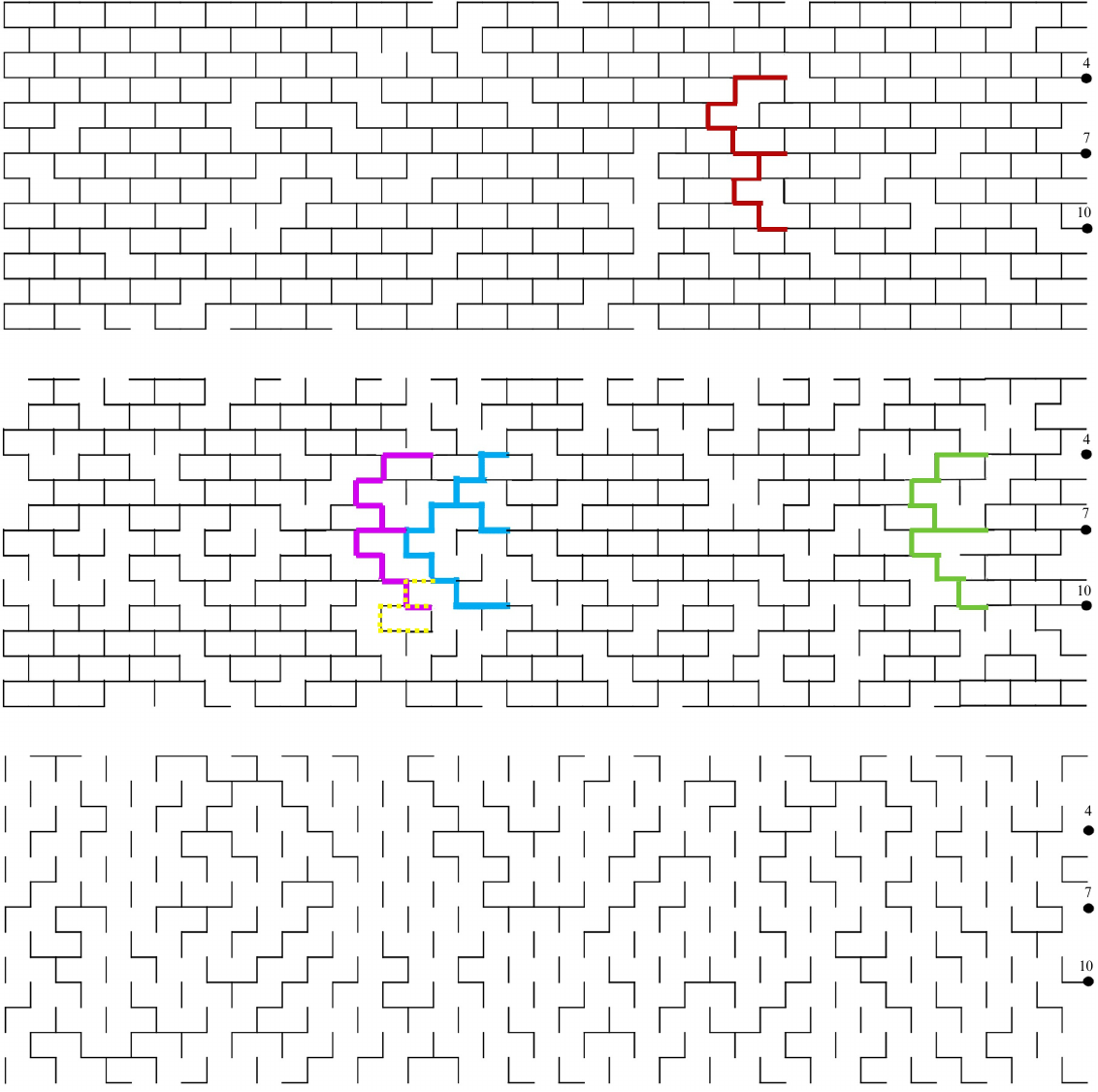}
\caption{{\bf Entanglement graphs}. Quantum circuits for the $L=14$ chain described in Fig.~\ref{fig:time}. Time increases from left to right. 
Vertical lines are unitaries, while horizontal lines denote the absence of measurement.
Top to bottom: $p=0.1,0.3,0.7$.
The 3 spins at positions $(4,7,10)$ composing $A$ are marked with black circles; their GME is discussed in the text and Fig.~\ref{fig:time}. 
Thicker colored bonds indicate \emph{minimal spanning graphs} which are necessary but not sufficient for GME. The dashed yellow lines in the middle panel are an example of a parasitic graph interfering with the purple minimal spanning graph.}
\label{fig:rootsL14}
\end{figure}

\section{Entanglement graphs} \label{sec_6}
\begin{figure}
\centering
\includegraphics[width=0.5\textwidth]{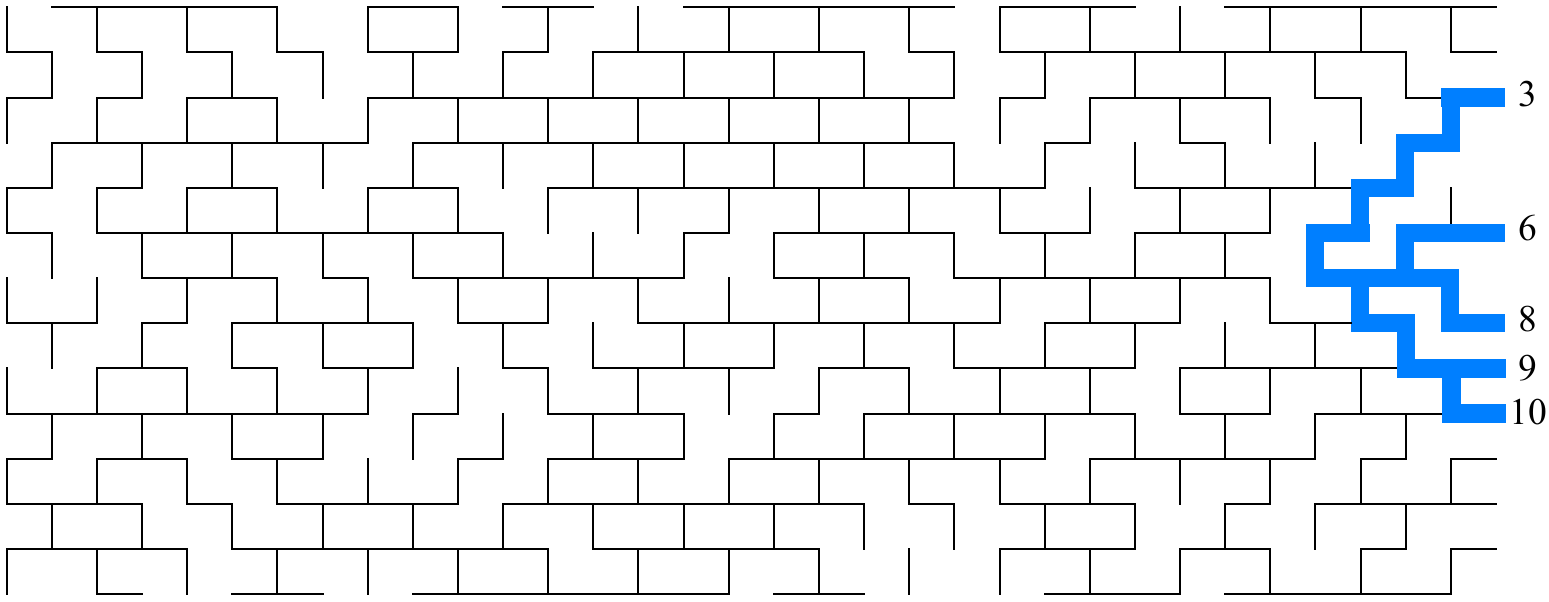}
\caption{{\bf Entanglement graph with 5 spins}. The time-evolution history for a $L=14$ circuit at $p=0.3$. The minimal spanning graph involving spins $(3,6,8,9,10)$ is shown.}
\label{fig:gme4}
\end{figure}

We wish to sharpen the graphical analysis introduced earlier to understand how GME spreads in complex monitored circuits. First, we need a more efficient representation that will allow us to tackle large systems for long times. We represent a 2-unitary by a vertical line joining the sites at equal time, and an absence of measurement by a horizontal line connecting a given site to the the next time step. This is illustrated in Fig.~\ref{fig:rootsL14}, with time increasing from left to right, for the $L=14$ circuits studied above. We want to understand the GME for a given subregion $A=A_1\cdots A_m$ ($B$ is the complement) by constructing certain graphs using the following rule:
\begin{itemize}
\setlength{\itemsep}{-3pt}
\item[$ $] Find the shortest graph connecting the largest number of spins in  subregions $A_1,\ldots, A_m$: we call this a \emph{minimal spanning graph} $G_{\rm min}$.
Examples are shown in Figs.~\ref{fig:rootsL14}-\ref{fig:gme4}. If $G_{\rm min}$ does not connect all subregions $A_1\cdots A_m$, $m$-party GME between the subregions cannot exist.


\end{itemize}
Overall, if $G_{min}$ has many connections with the graph composing the entire circuit, this weakens multipartite entanglement within $A$. In this respect, graphs that touch $A$ and $B$ can be called \emph{parasitic} as these tend to entangle $A$ with $B$, thus generically reducing the entanglement within $A$ (monogamy of entanglement). For example, a parasitic graph connecting spins 9, 10 and 11 interferes with the purple $G_{\rm min}$ in the middle panel of Fig.~\ref{fig:rootsL14} marked by a dashed yellow line. 
Furthermore, we find that shorter $G_{\rm min}$ graphs tend to produce stronger entanglement. 

We exemplify the construction and analysis of the minimal spanning graphs on the $L=14$ circuits. The bold purple and blue graphs at $p=0.3$ in the middle of Fig.~\ref{fig:rootsL14} represent $G_{\rm min}$ for the two
times at which GME has been detected (stars in Fig.~\ref{fig:time}). 
Nearby parasitic graphs that entangle $A$ with $B$ are limited, especially for the blue graph. In fact, the blue graph and its environment are nearly optimal.
Interestingly, the purple graph occurs again at a later time, as shown in green. However,  it has more parasitic graphs compared to the purple one, consistent with the fact that we detect no GME.
This generalizes to the low $p$ regime, as exemplified in the top circuit, where low depth $G_{\rm min}$ occur often, but they possess many connections with the rest of the circuit, including numerous parasitic graphs. 
An example is shown in red in the top circuit. In the case of high $p$, frequent measurements prevent the growth of graphs connecting distant spins, as is exemplified in the bottom circuit of Fig.~\ref{fig:rootsL14}, so that $G_{\rm min}$ does not connect all subregions $A=A_1\cdots A_m$ signalling the absence of $m$-party GME. 

The identification of $G_{\rm min}$ can be applied to general subregions. For instance, in the simpler case of 2 spins, we have evaluated  the logarithmic negativity  $\mathcal E(i,j)$ between sites $i,j\in\{4,7,10\}$ at both times where 3-spin GME has been detected (middle circuit of Fig.~\ref{fig:rootsL14}). For all choices of pairs, we found a non-zero  logarithmic negativity indicating bipartite entanglement. 
Furthremore, the above discussion leads to a prediction that $\mathcal E$ should be larger in the blue case compared to the purple one. We have found this to be true by a good margin for the three possible choices of pairs. 
As a more non-trivial example we have examined the 4-spin GME in another $L=14$ realization at $p=0.3$ shown in Fig.~\ref{fig:gme4}. 
We computed a measure of GME between 4 spins in the final state belonging to the subset $\{3,6,8,9,10\}$. These spins belong to a $G_{\rm min}$, and we have found strong 4-party GME between sites $\{6,8,9,10\}$ and $\{3,8,9,10\}$ using the $W_4$ criterion mentioned above. The range of 4-spin GME in the latter case is much larger than what is typically expected in equilibrium. 

Minimal spanning graphs are motivated by the observations made in the beginning of the paper, and are supported by our numerical calculations.
We have found that in the majority of instances where GME is found, a minimal spanning graph with a single seed connects the subregions. In these graphs one can follow the arrow of time from the seed to each of the subregions in $A$. However, less frequent instances exist where GME arises from a minimal spanning graph with multiple seeds. 

The entanglement graphs discussed above are classical objects in spacetime, and thus fit into the quantum-to-classical framework used to study various quantities in monitored quantum circuits~\cite{rev}. In particular, the graphs possess features in common with minimal cut and light cone structures~\cite{rev}, but are ultimately distinct. Further work is needed to understand the connection with percolation, and to determine whether entanglement graphs map to a quantitative statistical mechanical model. 

\section{Outlook}
We have studied how GME dynamically evolves in quantum circuits containing measurements and unitaries. We explained how an appropriate rate of measurements can lead to strong GME between distant subregions thus evading the usual fate of entanglement~\cite{Parez2024_F} in the scrambling regime. 
We have exemplified our general arguments with 1d random Haar circuits realizing a measurement-induced transition. We have found that 3-spin GME is strongest at intermediate measurement rates, and does not suffer a sudden death with separation.
We went further and found that bipartite and genuine tripartite entanglement \rd{are consistent with algebraical decay} with distinct quantum critical exponents, \rd{and pointed out the need for more careful analysis of finite-size and finite-ensemble effects}. These results shed light on the non-unitary (conformal) field theory that describes the transition, and further motivate the field-theoretic study of entanglement critical exponents.
We further examined the dynamics in specific realizations, and identified how long-range entanglement emerges. 

Finally, we have developed a graphical analysis based on minimal spanning graphs that connect the subregions. Such an approach allows one to see which spins are likely to share GME. We expect that it can be used to reveal further properties about quantum circuits. For example, in the Haar circuits under study, we see that for general rates $0<p<1$, nearly optimal minimal spanning graphs connecting arbitrary subregions can always be constructed, although often with low probability. This suggests that the ensemble average of a GME measure (such as $W$) should not suffer a sudden death with separation. 
The quantitative application of spanning entanglement graphs holds many promises, and future work is needed to reveal its full predictive power.

{\bf Acknowledgements---}We thank Y.~Hu, L.~Lyu, A.~Nahum, G.~Parez, J.~Pixeley, J.~Riddell and R.~Vasseur for useful discussions. W.W.-K.\ is supported by a grant from the Fondation Courtois, a Chair of the Institut Courtois, a Discovery Grant from NSERC, and a Canada Research
Chair. TPB is supported by NSERC and FRQNT grants. This research was enabled in part by support provided by the Digital Research Alliance of Canada (alliancecan.ca).

\clearpage
\onecolumngrid
\appendix

\begin{figure}
\centering
\includegraphics[width=0.6\textwidth]{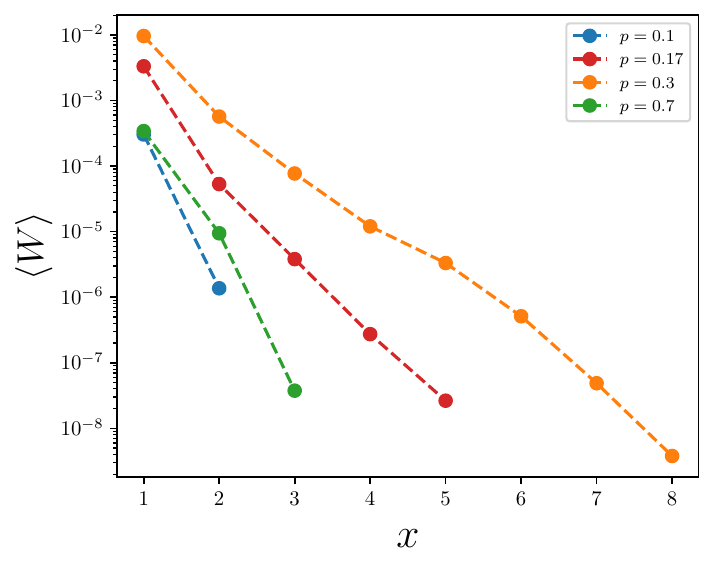}
\caption{{\bf 3-spin GME for $L=24$ \rd{with     open boundary conditions}}. Log-linear plot of the biseparability criterion $W$ for 3 spins at positions $(i,i+x,i+2x)$ for $L=24$ chains.
For $p=0.1,0.3,0.7$, the ensemble averages are taken over $10^4$ realizations while for $p=0.17$ the ensemble averages are over $2 \times 10^4$ realizations. For $p=0.1$,$0.17$, and $0.7$ positive $W$ events are detected for $x \leq 2$, $x \leq 3$, and $x \leq 5$ respectively while for $p=0.3$ events occur for $x \leq 8$. No positive events were detected for $x = 9,10,11$ due to the small number of realizations.}
\label{fig:L24}
\end{figure}


\section{Numerics}
\label{ap:numerics}
Calculations were in part performed using Qbit++: a general high-performance quantum many-body physics simulation engine \cite{qbitpp}. The circuit structure  comprises 4 layers that are repeated~\cite{Skinner2019,rev}. A first layer of unitaries is applied to the odd numbered bonds (the first consisting of the first two sites), then a measurement layer, another unitary layer on the even numbered bonds, and then another measurement layer. The unitary layers are comprised of random Haar matrices acting on two sites while the measurement layer is comprised of projective measurements in the $z$-basis. Each projective measurement occurs on each site with probability $p$. In Fig.~\ref{fig:W}, we apply a total of 98 layers, 49 unitary layers and 49 measurements layers. Note that we end with a unitary layer that covers every site (odd numbered bonds) and then a final measurement layer. The initial state is the all-0 product state in the computational basis.
To obtain $W$ for a given separation $x$ from the final state, we obtain the reduced density of 3 spins $\rho_{i_1,i_2,i_3}$ with $x=i_3-i_2=i_2-i_1$ and optimize over local unitary transformations to extremize $W$. This process is repeated for every possible $x$ in the chain. We then average over different circuit realizations in order to obtain $\langle W \rangle$ for all separations. 

\begin{figure}
\centering
\includegraphics[width=0.6\textwidth]{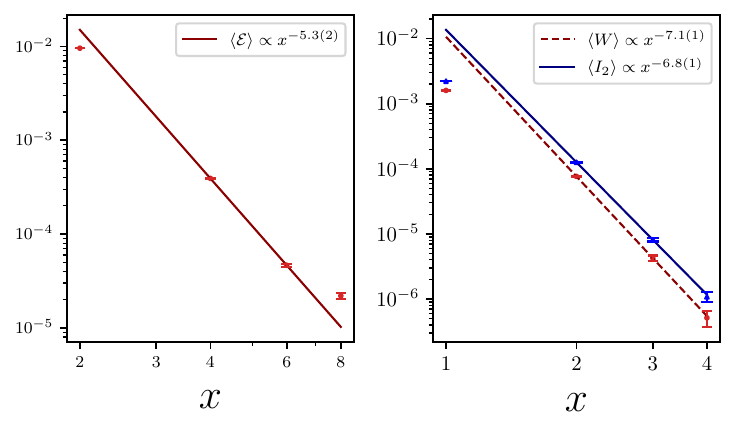}
\caption{\rd{{\bf{Scaling of $\langle \mathcal{E} \rangle, \langle W \rangle$, and $\langle I_{2} \rangle$ for $N=16$ near criticality}}. Log-log plot of the entanglement criteria versus separation $x$ for $N=16$ in periodic boundaries conditions at the rate $p=0.17$. The results are ensemble averaged over $1.17\times10^5$ realization for periodic boundary conditions with the corresponding error bars.}}
\label{fig:L16}
\end{figure}

\begin{figure}
\centering
\includegraphics[width=0.5\textwidth]{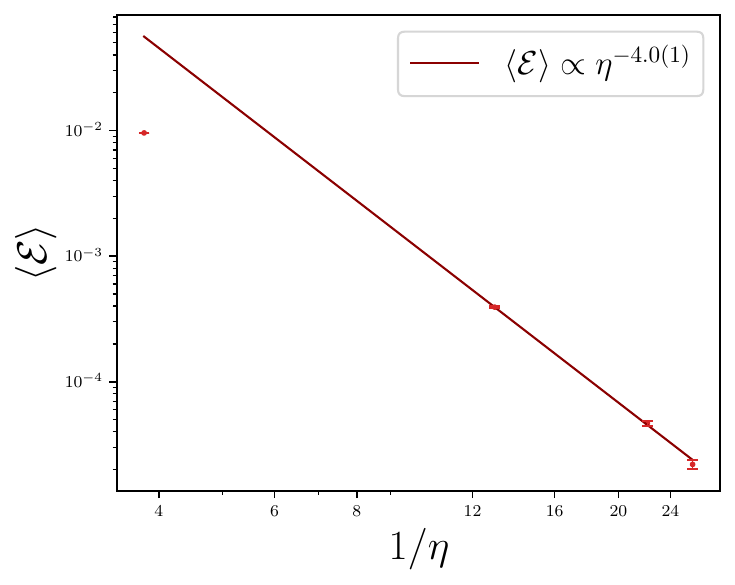}
\caption{\rd{{\bf{Scaling of $\langle \mathcal{E} \rangle$ near criticality vs CCR}}. Log-log plot of the entanglement criteria versus $1/\eta$ for $N=16$ in periodic boundaries conditions at the critical points $p_c=0.17$. The ensemble averaged logarithmic negativity is plotted for $x=2,4,6,8$.}}
\label{fig:L16_Chord}
\end{figure}

\begin{figure}
\centering
\includegraphics[width=0.5\textwidth]{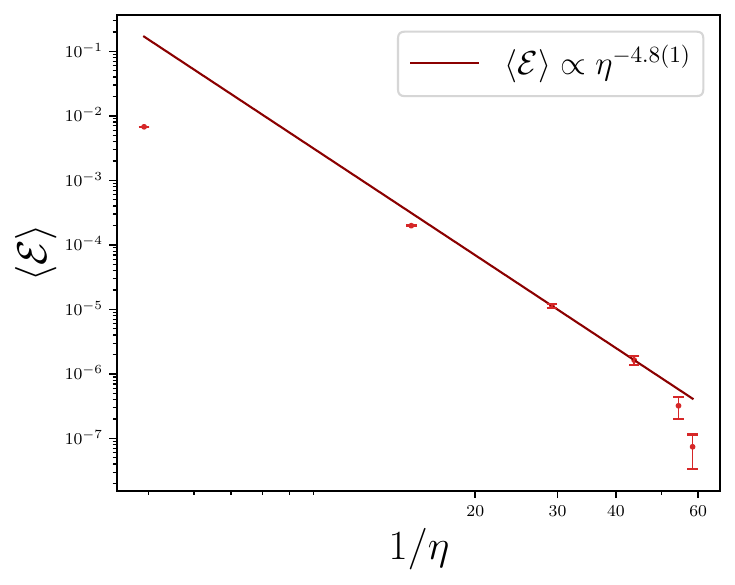}
\caption{\rd{{\bf{Scaling of $\langle \mathcal{E} \rangle$, at criticality vs CCR}}. Log-log plot of the entanglement criteria versus $1/\eta$ for $N=24$ in periodic boundaries conditions at the critical point $p_c=0.17$. The ensemble averaged logarithmic negativity is plotted for $x=2,4,6,8,10,12$ .}}
\label{fig:L24_Chord}
\end{figure}

In Figs.~\ref{fig:L24WPBC} and \ref{fig:L24NPBC}, we work with periodic boundary conditions to contrast with the results for the open boundary conditions. This further allows us to get closer to the $x$-dependence in the thermodynamic limit. Fig.~\ref{fig:L24NPBC} shows a simpler quantity: the logarithmic negativity between two sites separated by $x$ bonds. The error bars in Fig.~\ref{fig:powerlaw} are the standard deviation divided by the square root of the total number of samples used in calculating the mean from the $N=24$ data with periodic boundary conditions. \rd{The same treatment is repeated for Figs.~\ref{fig:L16}},~\ref{fig:L16_Chord},~\ref{fig:L24_Chord}

\rd{\subsection{Finite-size effects}}

\rd{To illustrate finite-size effects in Haar random circuits we show in Fig.~\ref{fig:L16} the $x$-dependence of the scaling exponents of $\langle \mathcal{E} \rangle$,$\langle W \rangle$, and $\langle I_2 \rangle$ at $N=16$ and $p=0.17$. We see that the exponents differ from  the results at $N=24$. For the logarithmic negativity we did not use the $x=2$ point in the fit as it is too small to be meaningful for the asymptotic scaling. We also ignored the $x=5$ as it shows an upturn due to the finite-size with periodic boundary conditions (below we discuss how to remedy this with the use of chord lengths). The exponent $\alpha_\mathcal{E}$ for $N=16$ indicates a slower decay of $\langle \mathcal{E} \rangle$ than for $N=24$, which can be understood as the shorter chain having fewer possible parasitic graphs that can impede the spreading of entanglement. For $W$ and $I_2$, the $x=5$ points are to be ignored in the fit as they are relatively rare, thus subject to large fluctuations in our finite ensembles. The $N=16$ data also has some non-zero hits at the maximal separation $x=5$ which is in contrast with the $N=24$ results which only have hits up to $x=5$ but has a maximum possible separation of $x=8$. Just like $\langle \mathcal{E} \rangle$, both $\langle W \rangle$ and $\langle I_2\rangle$ at $N=16$ decay slower than the $N=24$ results due to less parasitic graphs appearing in smaller chains. Moreover, the $N=16$ data respects the inequality of the exponents seen in the larger chain, namely that $\alpha_{\mathcal{E}} < \alpha_{W/I_2}$.} 

\rd{\subsection{Scaling of entanglement criteria using conformal cross ratio}}

\rd{We can analyze the scaling behavior of our entanglement criteria with respect to other kinds of distances to further discuss the validity of the exponents extracted. Prior studies have also looked at the conformal cross ratio (CCR) \cite{Sang2021,Shi2021}
\begin{equation} \label{eq:cr}
\eta = \frac{w_{12}w_{34}}{w_{13}w_{24}}\text{   }, \text{  }  w_{12} = \frac{N}{\pi}\sin\left(\frac{\pi}{N} |\rm{i}_1 - \rm{i}_2|\right)  
\end{equation}
for chains with periodic conditions and where $w_{12}$ is the chord length between the sites on the dual lattice $\rm{i}_1$ and $\rm{i}_2$ that bound the first subregion  (similarly for $w_{34}$ which corresponds to the second subregion, and the other separations). The CCR is typically used to study subregions with multiple sites, but can also be used for subregions containing only a single site. In such a case, the CCR reduces to 
\begin{equation}
\eta = \frac{1}{w^2_{ij}} = \frac{1}{(\frac{N}{\pi}\sin(\frac{\pi}{N}x))^2}
\end{equation} 
where $x$ is the bond distance defined in the main text. Moreover, this choice of distance compared to the bond distance may be more suitable to the geometry of the circular lattice when trying to study maximal separations. 
Another reason to use this notion of distance is that the CCR appears as the argument of a scaling function that exhibits conformal symmetry \cite{Li2019}. 
When trying to extract a scaling exponent of a quantity with $\eta$, the discrete nature of the lattice and the small number of unique separations between subregions makes it difficult to see the asymptotic scaling. It is to be noted that if intervals are used rather than single spin regions, there are even less separations possible that can be used to determine the scaling. This can be seen in Figs.~\ref{fig:L16_Chord} and \ref{fig:L24_Chord}, where the larger $x$ increases, only few points have large enough separations that meaningfully contribute to the fit to the asymptotic form $\eta^\alpha$ while also having enough non-zero hits from the ensemble average. For $\langle \mathcal{E} \rangle$ (Fig. ~\ref{fig:L16_Chord}), the $N=16$ scaling looks acceptable but the separations in the chain are too small to infer the validity of the scaling exponent. When going to the $N=24$ case (Fig. ~\ref{fig:L24_Chord}), the added points suffer from a rarity of entangling events at the larger separations. The $\langle W \rangle$ and $\langle I_2 \rangle$ suffer the same issues when plotting with the CCR as they do versus the bond distance. Namely, the detection of entanglement with imperfect measures does not improve by changing our notion of distance. 
}

\section{Four-party genuine entanglement}
\label{ap:four}
We discuss examples of GME involving 4 spins obtained in the final state of the $L=14$ circuit shown in Fig.~\ref{fig:gme4}. A general 4-spin state is given by a 16-by-16 density matrix with elements $\rho_{ij}$, $1\leq i,j\leq 16$, where we use the standard computational basis,
$\{|0\cdots 0\rangle, |0\cdots 10\rangle,\dots, |1\cdots 1\rangle\}$. 
The bi-separability criterion that we shall use is a close cousin of the 3-spin criterion given in the main text. It reads \cite{W_criterion}
\begin{multline} \label{eq:w4}
W_4= \max_{\rm LF}| \rho_{2,3}| + |\rho_{2,5}| + |\rho_{2,9}| + |\rho_{3,5}| + |\rho_{3,9}| + |\rho_{5,9}|
   -\rho_{2,2} -\rho_{3,3} -\rho_{5,5} -\rho_{9,9}\\
   -\sqrt{\rho_{1,1} \rho_{4,4}}
   -\sqrt{\rho_{1,1} \rho_{6,6}}
   -\sqrt{\rho_{1,1} \rho_{7,7}}
   -\sqrt{\rho_{1,1} \rho_{10,10}}
   -\sqrt{\rho_{1,1} \rho_{11,11}}
   -\sqrt{\rho_{1,1} \rho_{13,13}}
\end{multline}
where the maximisation is over all \emph{local filter} (LF) operations, $\rho\mapsto  (F_1\otimes F_2\otimes F_3\otimes F_4) \rho (F_1^\dag \otimes F_2^\dag \otimes F_3^\dag\otimes F_4^\dag)$, where the $F_i$ are arbitrary 2-by-2 matrices. 
The set LF includes local unitaries as a small proper subset, and is thus more powerful. When $W_4>0$, we conclude that the system has 4-party GME. If $W_4=0$ (within machine precision), we cannot conclude anything, and other methods need to be used. Note that $W_4<0$ is not possible since LF can always bring the state into the zero matrix.
This criterion was obtained from the 4-spin $W$ state $|W_4\rangle =  \tfrac{1}{\sqrt 2}(|0001\rangle+|0010\rangle+|0100\rangle+|1000\rangle)$. For instance, the first 6 terms in Eq.~\eqref{eq:w4} are the norms of the off-diagonal elements of $\rho$ in the upper triangle that correspond to the non-vanishing matrix entries of $|W_4\rangle \langle W_4|$.


\newpage
\newpage
\twocolumngrid
\bibliography{references}

\end{document}